\documentclass[12pt,letterpaper]{iopart}

\usepackage{charter,graphicx,amstext}

\newcommand{\aro}{$A_2$Ru$_2$O$_7$}
\newcommand{\aaro}{$A_3$RuO$_7$}
\newcommand{\abro}{$A_{2-x}$Bi$_x$Ru$_2$O$_7$}
\newcommand{\bro}{Bi$_2$Ru$_2$O$_7$}
\newcommand{\bto}{Bi$_2$Ti$_2$O$_7$}
\newcommand{\yro}{Y$_2$Ru$_2$O$_7$}
\newcommand{\nro}{Nd$_2$Ru$_2$O$_7$}
\newcommand{\nnro}{Nd$_3$RuO$_7$}
\newcommand{\degree}{$^{\circ}$}
\newcommand{\Rwf}{$R_{\text{wf}}$}
\newcommand{\Rwp}{$R_{\text{wp}}$}
\begin{document}

\title[Structural disorder, magnetism, and thermoelectricity of \nro\/]
{Structural disorder, magnetism, and electrical and thermoelectric 
properties of pyrochlore \nro\/}

\author{Michael W Gaultois$^1$, Phillip T Barton$^1$, Christina S Birkel$^1$,\\ 
Lauren M Misch$^1$, Efrain E Rodriguez$^2$, Galen D Stucky$^1$ \\ and Ram Seshadri$^1$} 
\address{$^1$ Materials Department, Department of Chemistry and Biochemistry\\
and Materials Research Laboratory, 
University of California, Santa Barbara, CA 93106, USA}
\address{$^2$ Department of Chemistry and Biochemistry\\
University of Maryland, College Park, MD 20742, USA}
\eads{mgaultois@mrl.ucsb.edu}

\begin{abstract} Polycrystalline \nro\/ samples have been 
prepared and examined using a combination of structural, magnetic, and 
electrical and thermal transport studies. Analysis of synchrotron X-ray 
and neutron diffraction patterns suggests some site disorder on the 
\textit{A}-site in the pyrochlore sublattice: Ru substitutes on the Nd-site 
up to 7.0(3)\%, regardless of the different preparative conditions explored. 
Intrinsic magnetic and electrical transport properties have been measured. 
Ru 4d spins order antiferromagnetically at 143\,K as seen both in susceptibility
and specific heat, and there is a corresponding change in the electrical 
resistivity behaviour. A second antiferromagnetic ordering transition 
seen below 10\,K is attributed to ordering of Nd 4f spins. \nro\/ is
an electrical insulator, and this behaviour is believed to be
independent of the Ru-antisite disorder on the Nd site. The electrical
properties of \nro\/ are presented in the light of data published on all
\aro\/ pyrochlores, and we emphasize the special structural role that Bi$^{3+}$
ions on the \textit{A}-site play in driving metallic behaviour. 
High-temperature thermoelectric properties have also been measured. 
When considered in the context of known thermoelectric materials with 
useful figures-of-merit, it is clear that \nro\/ has excessively high 
electrical resistivity which prevents it from being an effective 
thermoelectric. A method for screening candidate thermoelectrics
is suggested. 

\pacs{75.47.Lx 
      75.50.Ee 
      75.50.Lk 
      72.20.Pa 
     }
\end{abstract}

\maketitle

\section{Introduction} 

\begin{figure}
\centering 
\includegraphics[width=0.75\textwidth]{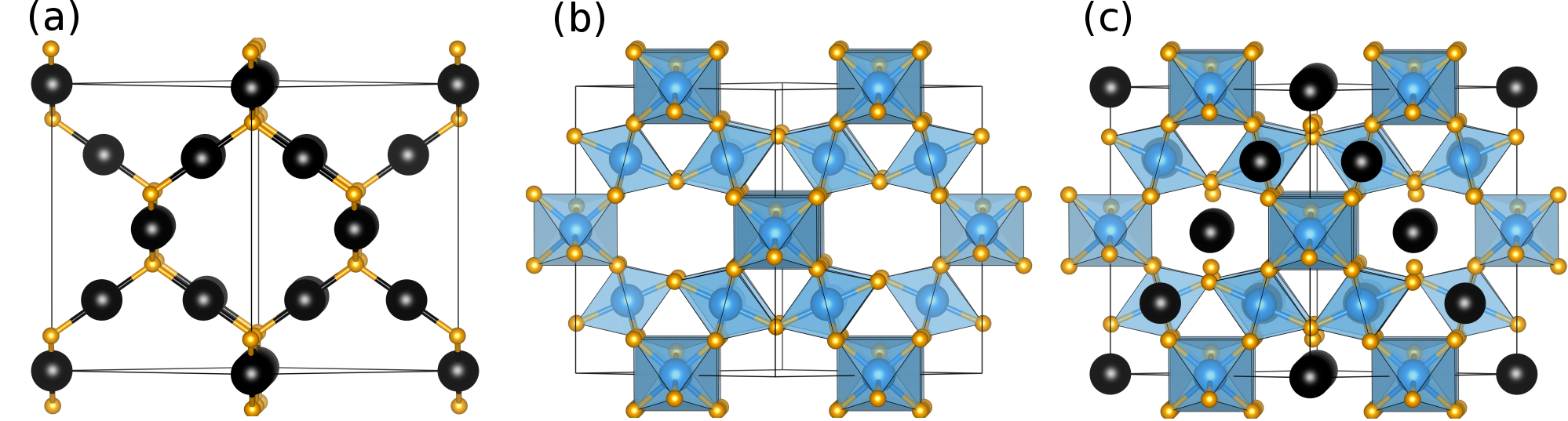} 
\caption{Interpenetrating a (a) ``cubic-ice'' diamondoid $A_2$O$^\prime$ 
lattice with (b) a lattice of corner-connected $B$O$_{6}$ octahedra
gives the  (c) $A_2B_2$O$_6$O$^\prime$ pyrochlore structure of \nro.
Space group $Fd\bar3m$ [No. 227, \textit{A} at $(\frac12,\frac12,\frac12)$, 
\textit{B} at $(0,0,0)$, O at $(x_O,\frac18,\frac18)$, and O$^\prime$ at 
$(\frac38,\frac38,\frac38)$]. Black spheres are Nd, blue octahedra surround 
Ru, and the orange spheres are O.}
\label{fig:structure}
\end{figure}

The $A_{2}B_{2}$O$_6$O$^\prime$ pyrochlore structure, shown
in figure\,\ref{fig:structure}, comprises two interpenetrating
$B_2$O$_6$ and $A_2$O$^\prime$ sublattices \cite{Subramanian1983PSSC}. 
The electrical properties of pyrochlore ruthenium oxides (\aro\/, 
\textit{A}= Pr through Lu, Y, and Bi) are of long-standing interest. 
In \aro\/ ruthenates, electrical conductivity is expected to take place 
through the Ru$_2$O$_6$ network of RuO$_6$ octahedra, \textit{ie.} the 
$B_2$O$_6$ sublattice. While all the rare-earth members are insulating, 
\bro\/ is metallic. Several investigators have sought to explain the 
difference in conductivity by examining changes in Ru--O--Ru bond angle 
--- determined by the combination of Ru--O bond length and the size of 
\textit{A} --- and consequent changes in orbital overlap and 
bandwidth \cite{Lee1997JSSC,Li2003CM}. Electron spectroscopic 
investigations of \yro\/ compared with \bro\/ led 
Cox \textit{et al.}\cite{Cox1983JPCSSP} to conclude that the principle 
difference seen is the participation in conduction of Bi 6s states in \bro\/. 
This participation broadens the Ru 4d band width sufficiently that a metallic ground 
state is preferred over one that is correlated and insulating. However,
Shoemaker \textit{et al.}\cite{Shoemaker2011PRB} contrasted the 
computed electronic structures of insulating \bto\/ with that of 
conducting \bro\/ and found no difference in the presence or
absence of Bi 6s states near the Fermi energy in these two compounds.

Pyrochlore-type rare-earth ruthenium oxides are also of great interest for
their magnetic properties. The sub-lattices of corner-connected
$A_4$ and $B_4$ tetrahedra can result in magnetic frustration when either 
\textit{A} or \textit{B} is separately magnetic \cite{Gardner2010RMP}. While magnetic 
properties of \nro\/ have been reported previously \cite{Taira1999JPCM}, we 
show here that the presence of ferromagnetic impurity phases compromised the 
earlier analysis. There has also been some confusion in the 
literature about the nature of the ordering of the Ru 4d spins in \nro\/; 
several reports have suggested that the ordering is 
glassy \cite{Taira2000JSSC,Ito2000JPSJ,Ito2001JPCS,Kmiec2006PRB,Gurgul2007PRB}.
We find compelling evidence that \nro\/ is not a spin-glass, and propose
a reason for the observed history-dependence of the ZFC-FC magnetic 
susceptibility and weak ferromagnetism below the ordering temperature. 
Heat capacity studies of the magnetic transition in \nro\/ are also presented 
here. 

In \aro\/ pyrochlores the structural modifications between metallic and 
semiconducting members are small, suggesting that the electrical transport 
properties are delicately positioned near the edge of localized and itinerant 
behaviour. This positioning, at the edge of metal-insulator divide, potentially
serves as a prime locator for thermoelectric materials: a high Seebeck 
coefficient is generally found in insulators, with potentially acceptable 
electrical conductivity on the metallic side, striking the right balance in 
properties \cite{Snyder2008NM,Tritt2006MB}. For this reason, we investigate 
the high-temperature thermoelectric properties of \nro. When compared to 
well-known chalcogenide thermoelectric materials, for \textit{e.g.}, PbTe, 
Bi$_2$Se$_3$, and Bi$_2$Te$_3$, metal oxides are of interest due to the 
expectation of higher stability at elevated temperatures and the prospect 
of using less toxic and more abundant elements. The discovery of high 
thermoelectric performance in Na$_x$CoO$_4$\cite{Terasaki1997PRB} has led 
to renewed interest in oxide thermoelectric materials, despite the figure 
of merit (\textit{ZT}) of this system and other bulk oxide materials being too small 
for widespread use \cite{Tritt2006MB,He2011JMR}. Towards the goal of a more 
directed and effective search for high-performance metal-oxide thermoelectric 
materials, we also introduce a new type of plot for data visualization that can be 
used to rapidly screen potential candidates for thermoelectric performance.

\section{Experimental details}

Polycrystalline samples were made by direct reaction of constituent oxide 
powders (RuO$_{2}$, 99.99\%, Sigma-Aldrich; Nd$_{2}$O$_{3}$, 99.99\%, 
Alfa Aesar). Owing to the volatility of Ru oxides in air
at high temperatures ($\geq$1040\degree C), samples were prepared with a
1\,mol\% excess RuO$_{2}$. Pellets were cold-pressed and annealed at
1000\degree C for 7 days with several intermediate grindings, after which 
starting materials were still found by lab XRD. Following this
initial reaction, pellets were wrapped in Pt-foil and annealed in evacuated
silica ampoules at 1100\degree C for 7 days. The samples were
cooled to 800\degree C at 0.5\degree C/min, and further annealed
at 800\degree C for 7 days to promote healing of defects. Finally, samples were
cooled slowly to room temperature at 0.5\degree
C/min.  If \nro\/ is annealed in air at 1040\degree C or 1060\degree C 
for 12 days, Rietveld analysis of XRD patterns indicates the sample contains
78\,mol\% \nnro\/ and 22\,mol\% \nro\/ when the annealing temperature is
1040\degree C, and pure \nnro\/ when the annealing temperature is 1060\degree C.

To avoid the formation of \nnro\/ at the high temperatures and extended
annealing times required for solid state reactions, polycrystalline samples
of $>$99\% purity were also made by ultrasonic spray pyrolysis (USP) of
aqueous nitrate precursors [Ru(NO)(NO$_3$)$_3$, Sigma-Aldrich; and an aqueous
solution of Nd$^{3+}$, prepared by dissolving Nd$_{2}$O$_{3}$ in $\approx$2\,M 
HNO$_{3}$]\cite{Skrabalak2006JACS,Misch2011CM}. A modified ultrasonic humidifier was used 
to generate a mist, which was passed through a tube furnace (30.5\,cm heating 
zone) at 700\degree C with a positive pressure of air at a flow-rate of 
5\,scfm ($\approx$140\,L/min).  The product was collected
in water, and the mixture was evaporated at 70\degree C; the resulting black
solid was cold-pressed into a pellet and annealed in air at 1100\degree C for
8 hours. Samples were then quenched in air (removed from the furnace at
1100\degree C) or annealed in air for 7 days at 800\degree C then cooled to
room temperature at 1\degree C/min. Rietveld refinement of the structure with synchrotron XRD
data revealed no significant differences in the bulk long-range structure
between samples that were quenched rapidly and samples that were annealed at
intermediate temperatures and cooled slowly, nor were there differences in
structure between samples made by USP and samples made by solid state
reaction. USP offers a rapid preparatory route to prepare \nro\/ in $\approx$24
hours, rather than $\approx$14 days.

Laboratory XRD was performed using a Philips X'Pert diffractometer with 
Cu K$\alpha$ radiation and using mis-cut Si sample holders to reduce background 
signal. High-resolution synchrotron X-ray diffraction (XRD) data on 
finely-ground powder was acquired at 100\,K and 295\,K at beamline 11-BM at 
the Advanced Photon Source (APS), Argonne National Laboratory, using an 
average wavelength of 0.413\,\AA\/ on a diffractometer that has been
described in detail by Wang \textit{et al.} \cite{Wang2008RSI}. 
The precise wavelength was determined using a mixture of Si (SRM 640c)
and Al$_{2}$O$_{3}$ (SRM 676) NIST standards run in a separate 
measurement. Samples were contained in 0.4\,mm diameter Kapton capillaries 
and the packing density was low enough that absorption was not noticeable.  
Neutron powder diffraction was performed on the BT-1 
diffractometer at the NIST Center for Neutron Research. 
A Cu(311) monochromator was used, with a constant wavelength of 
$\lambda$ = 1.5402(2)\,\AA\/ and a second-order contribution at 
$\lambda/2$. Data was collected at 300\,K over the range of 
3$^{\circ}$ to 168$^{\circ}$ 2$\theta$ with a step size of 0.05$^\circ$. 
All diffraction data shown here are from a sample made by solid state 
reaction, though Rietveld analysis of data obtained by synchrotron XRD on 
other samples made by solid state reaction or USP lead to identical results.

X-ray total scattering was performed at beamline 11-ID-B at the APS, using a
wavelength of 0.137020\,\AA\/. The pair distribution function (PDF) was
extracted with PDFgetX2 \cite{Qiu2004JAC} using
$Q_{max}$ = 28\,\AA\/$^{-1}$, and full-profile PDF structure refinement
was completed using PDFgui \cite{Farrow2007JPCM}. \Rwf\/ and $\chi^2_{red}$
were refined on 1950 data points. Key instrumental 
parameters were $Q_{broad}$ = 0.0551\,\AA\/$^{-1}$ and $Q_{damp}$ = 
0.00963\,\AA\/$^{-1}$ respectively, using a CeO$_{2}$ standard run in a 
separate measurement.

Magnetic properties of powders were measured using a Quantum Design MPMS XL-5
SQUID magnetometer. In addition to DC measurements, frequency-dependent AC 
measurements were performed in a small temperature range between 130\,K and 
155\,K. Low-temperature electrical transport properties and heat capacity were
measured using a Quantum Design Physical Properties Measurement System.
Samples for electrical transport measurements employed the 4-probe geometry on 
a pellet of sintered powder with dimensions of approximately 9\,mm$ \times$ 
3\,mm $\times$ 3\,mm. Electrical contacts were made with copper
wire and silver epoxy. Three samples were run to ensure reproducibility: two 
made using USP with apparent densities of 38\% and 53\%, and one made by 
ceramic preparation with an apparent density of 58\%. Electrical resistivity 
at 300\,K varied from 1.2\,$\Omega$\,cm to 2.4\,$\Omega$\,cm between the 
samples, and is consistent with previous reports of 1.8\,$\Omega$\,cm 
\cite{Yamamoto1994JSSC}. Despite slight differences in magnitude of 
resistivity, all
samples displayed consistent temperature-dependent behaviour. Heat capacity
measurements were collected on a pellet of mass 13.90\,mg made by pressing
50\,wt\% of sample and 50\,wt\% of silver powder (99.99\%, Sigma-Aldrich)
and analyzed using the thermal relaxation dual-slope method. 
A thin layer of Apiezon N grease ensured thermal contact between 
the platform and the sample. The heat capacity of the Apiezon N grease and 
silver were collected separately and subtracted from the measured sample heat 
capacity.

High-temperature thermoelectric properties (electrical resistivity and
Seebeck coefficient) were measured with an ULVAC Technologies ZEM-3.
Sample pellets had approximate dimensions of 9\,mm$\times$3\,mm$\times$3\,mm.
 Measurements were 
performed under a helium under-pressure, and data was collected through three 
heating and cooling cycles to ensure sample stability and reproducibility. No
changes in physical properties were observed between cycles, and analysis of
the lab XRD pattern of the materials after measurements showed no changes in
structure nor new phases. Two different samples (USP and ceramic) were tested 
to verify consistency.

\section{Results and Discussion}

\subsection{Structure}

The high symmetry ideal pyrochlore crystal structure is completely determined
by the cubic cell parameter and a single positional structural parameter,
associated by the position $x$ of O which is sited at $(x_O,\frac18,\frac18)$. 
However, there are many types of disorder that arise in this structure type, 
and a careful examination of the structure is necessary to understand 
properties. For example, Vanderah \textit{et al.} recently described the 
widespread presence of antisite substitution on the \textit{A}-site; up to 
25\% of the large \textit{A}-sites can be replaced with small \textit{B}-site 
metal ions \cite{Vanderah2005EJIC}. Moreover, local off-centering of 
\textit{A}-site cations is well-known in Ru pyrochlores such as \bro\/ 
\cite{Shoemaker2011PRB,Avdeev2002JSSC}. The 
presence of disordered oxygen vacancies has been observed in metallic members 
of \abro\/ solid solutions \cite{Kennedy1996JSSC}. Due to the relatively 
small X-ray scattering factor of O compared to the other elements present 
in \aro\/, use of only lab XRD has led to inaccurate determination of the 
O atomic position, as demonstrated by Kennedy and Vogt \cite{Kennedy1996JSSC}. 
This inaccuracy has a large effect on the reported Ru--O--Ru bond angles, 
which are known to critically influence electrical properties. For example, 
in \bro\/, there is a 6\degree\/ discrepancy, and for other \aro\/ members 
there is up to a 2\degree\/ discrepancy in the reported Ru--O--Ru bond angles
\cite{Li2003CM,Avdeev2002JSSC,Kanno1993JSSC}. 

\begin{figure}
\centering 
\includegraphics[width=0.5\textwidth]{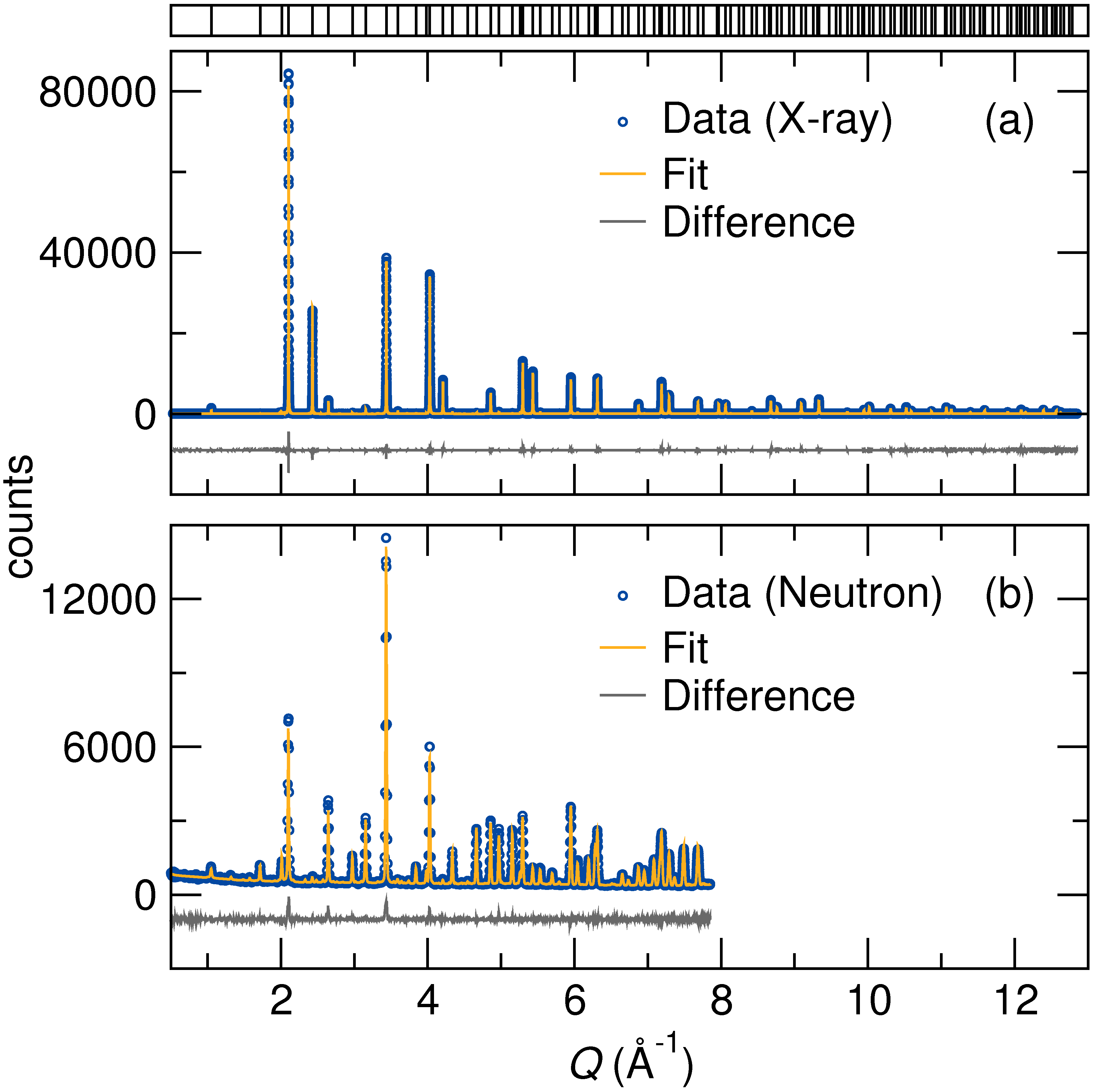} 
\caption{Combined Rietveld refinement of the structure with 
(a) room-temperature synchrotron XRD with $\lambda=0.41295$\,\AA\/ and
(b) neutron diffraction with $\lambda=1.5402$\,\AA\/. From the refinement,
the cell parameter $a$ is determined to be 10.342312(8)\,\AA\/ at room 
temperature, and the position $x_O$ corresponding to the O atom is 0.33012(7).}
\label{fig:Rietveld}
\end{figure}

Combined Rietveld refinement was carried out using 
room-temperature data sets (figure\,\ref{fig:Rietveld}), where the structure
was refined on 51295 data points. During the combined 
refinement, the synchrotron X-ray wavelength was fixed while the neutron 
wavelength was allowed to vary, though the refined wavelength was within 
two standard deviations of the starting value determined by previous 
instrumental calibration. Isotropic displacement parameters were used to 
describe the electron density of the atoms in the structure, as refinement 
of anisotropic displacement parameters did not improve the quality of the fit. 
Rietveld refinement was performed with XND code \cite{Baldinozzi1997JPCM}, and 
structures were visualized using VESTA \cite{Momma2011JAC}. 

Rietveld refinement of the structure using synchrotron powder XRD (figure\,\ref{fig:Rietveld}(a))
data indicates samples were 99.5(1)\,mol\% \nro\/ and 0.5(1)\,mol\% RuO$_2$. 
The minor presence of RuO$_2$ is not expected to influence the physical properties reported here.
There were no significant differences in structure between samples that 
were quenched rapidly and samples that were annealed at intermediate 
temperatures and cooled slowly. Additionally, there were no significant 
differences in structure between samples made by USP and samples made by solid 
state reaction, nor were there differences between samples annealed under 
static vacuum and samples annealed in air. Taken together, the long-range 
structural order of \nro\/ appears to be insensitive to preparation conditions 
and methods. In particular, annealing \nro\/ under low oxygen partial pressures
(\textit{i.e.}, in evacuated ampoules) does not appear to lead to oxygen
deficiency. More evidence of this stability is presented in a later subsection,
as electrical resistivity does not change after several heating cycles between 
300\,K and 900\,K under oxygen-free conditions.
This suggests that O vacancies not already present in the ordered pyrochlore
structure are not formed to an appreciable extent in \nro\/.
To check for the presence of oxygen vacancies, the site occupancy of 
O$^\prime$ was allowed to refine freely; the best refinements were
consistent with complete occupancy. This finding is supported by previous
neutron-diffraction studies, which found no evidence for additional oxygen
vacancies in \nro\/ \cite{Kennedy1996JSSC}.
The oxygen positional parameter $(x_O,\frac18,\frac18)$ converged 
to $x_O$ = 0.33012(7), identical to previous reports using neutron diffraction 
\cite{Kennedy1996JSSC,Field2000JSSC}. 

\begin{figure}
\centering 
\includegraphics[width=0.5\textwidth]{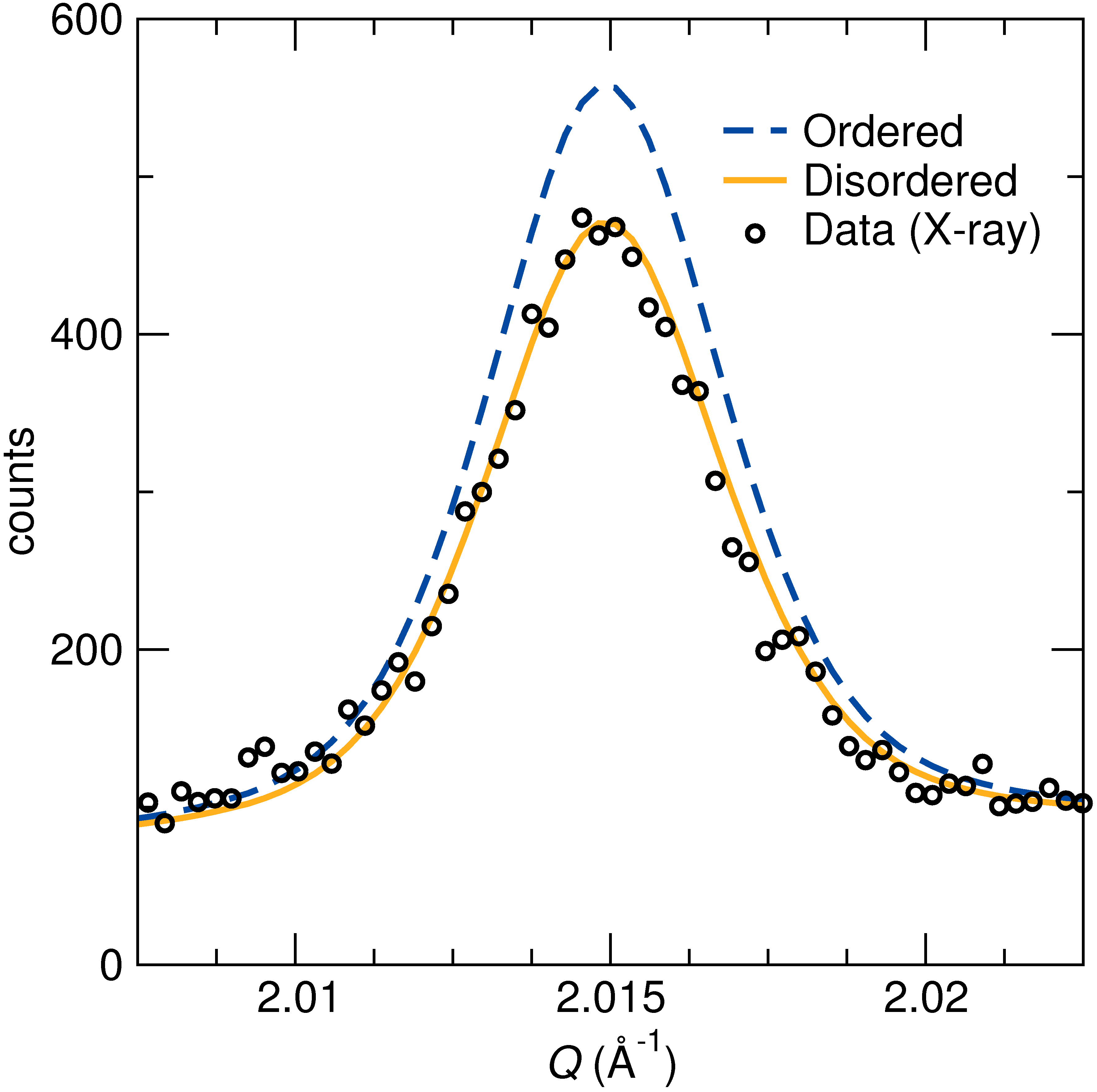} 
\caption{Refinement with \textit{A}-site disorder converges to 7.0(3)\% Ru on
the \textit{A}-site and slightly improves the fit to the data, as
demonstrated above by the better description of the peak-shape. Changes in
the calculated diffraction pattern due to site disorder are small, with the
most significant and diagnostic change occurring at the 311 reflection. 
For comparison, the most intense peak in this pattern has 85\,000 counts.}
\label{fig:antisite_disorder}
\end{figure}

Synchrotron X-ray diffraction studies offer the advantage that the Nd$^{3+}$ 
and Ru$^{4+}$ X-ray scattering factors are sufficiently distinct due to the 
large difference in atomic numbers. In contrast, neutron diffraction does
not clearly distinguish between the similar coherent scattering lengths of 
Nd (7.03\,fm) and Ru (7.69\,fm) \cite{Sears1992NN}. The synchrotron 
diffraction data allowed the recent findings of Vanderah \textit{et al.} 
regarding antisite disorder on the cation sites to be tested. The site 
occupancy and atomic displacement parameters are often correlated, so 
synchrotron X-ray data was collected at both 100\,K and room temperature 
(295\,K). However, these parameters were not strongly correlated in this 
investigation, and a combined Rietveld refinement using multiple temperatures 
did not change the outcome of the analysis. Accordingly, we continue our 
discussion using the combined Rietveld refinement of room-temperature X-ray 
and neutron diffraction datasets (figure\,\ref{fig:Rietveld}). 
When Ru was allowed to substitute on the 
\textit{A}-site, the refinement converged with 7.0(3) mol\% Ru and a slight 
improvement in the fit. A similar trial refinement of Nd on the 
\textit{B}-site did not improve the fit. The stability of all refined models 
(\textit{B}-site disorder, \textit{A}-site disorder, no antisite disorder) was 
verified by perturbing other parameters. Because allowing \textit{A}-site 
disorder improves the fit only marginally, it is important to consider whether 
the improved fit is significant, or if the improvement is merely because more 
parameters are introduced. Use of a Hamilton test \cite{Hamilton1965AC} shows 
the difference is statistically significant at $<$0.5\% confidence interval. 
Indeed, with only one additional parameter between the two models, the large 
number of independent measurements makes virtually any improvement in \Rwp\/ 
statistically significant. Visual inspection reveals only minor changes 
between the models, though in the limited areas where antisite disorder causes 
the largest changes, \textit{A}-site disorder improves the fit to the 
experimental peak-shape (figure\,\ref{fig:antisite_disorder}). The same 
results are obtained in samples made by USP and by solid state reaction. With 
these considerations in mind, the presence of \textit{A}-site disorder 
is suggested.

\begin{figure}
\centering 
\includegraphics[width=0.5\textwidth]{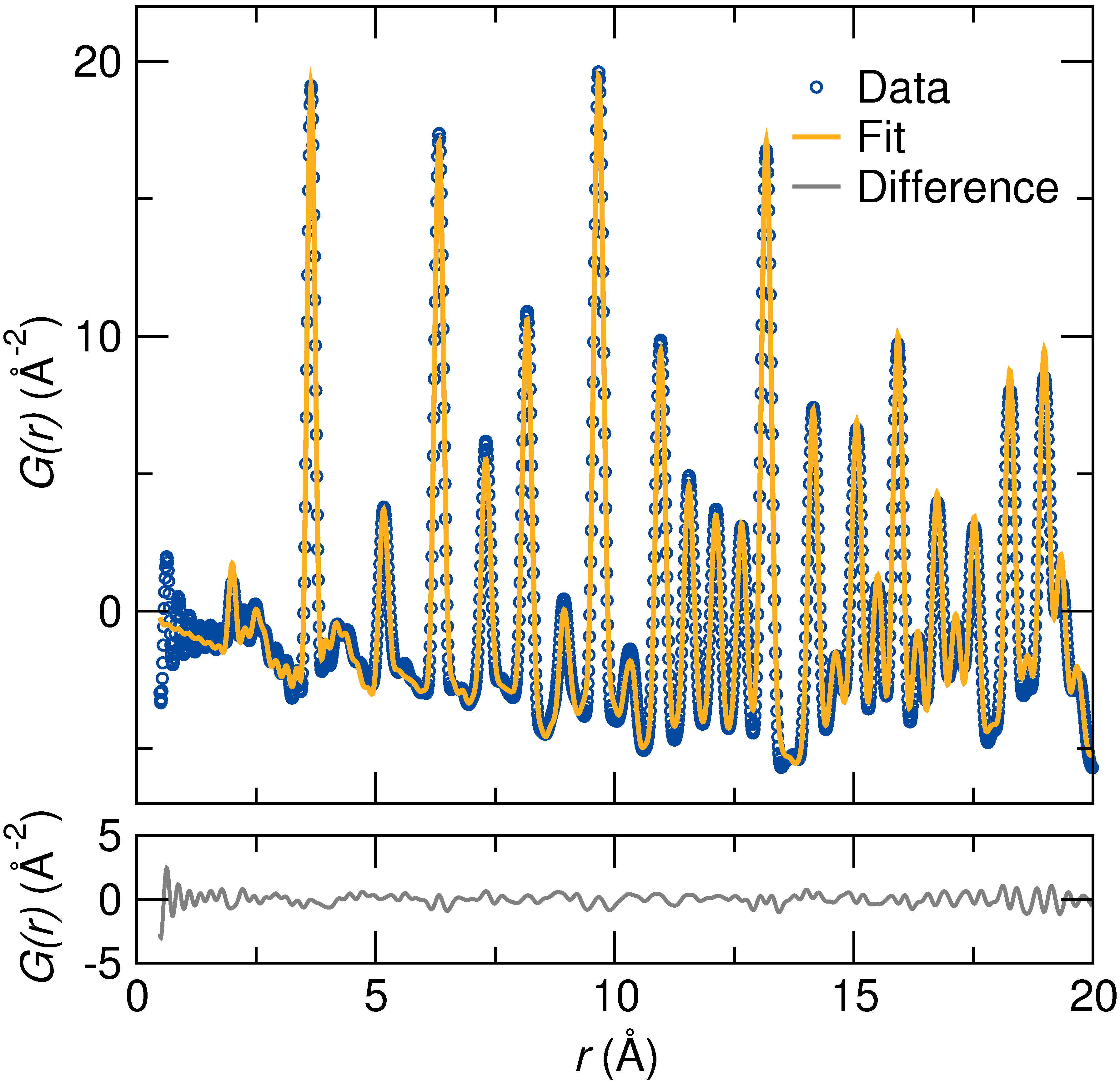} 
\caption{Analysis of the synchrotron X-ray PDF reveals \nro\/ is
well-described by an average long-range structural model.}
\label{fig:xpdf}
\end{figure}

The X-ray pair distribution function (PDF) obtained by total scattering
agrees well with the model generated by the average, long-range structure
(figure\,\ref{fig:xpdf}).  Refinement of partial Ru substitution on the
\textit{A}-site was attempted, but the refined occupancy converged to 
unphysical values, potentially due to strong correlation with the scale
factor. Ideal cation site ordering \nro\/ yielded a refinement $R$ = 10.05\%
and allowing 7.0\% Ru to occupy the Nd site, as suggested from Rietveld 
refinement, yielded $R$ = 10.02\%. The numerical improvement of the fit is 
marginal, and visual inspection reveals the difference between the models is 
much less than the level of noise present in the fit to the data.
Rietveld refinement of Bragg scattering suggests partial antisite disorder, 
whereas PDF analysis of the total scattering shows no strong preference 
between full ordering or partial antisite disorder. This likely occurs because 
the PDF refinement is strongly weighted by near-neighbor correlations, which 
has been previously noted in the system La$_4$LiAuO$_8$ \cite{Kurzman2010IC}.

\subsection{Magnetic and electrical transport behaviour}

\begin{figure}
\centering 
\includegraphics[width=0.5\textwidth]{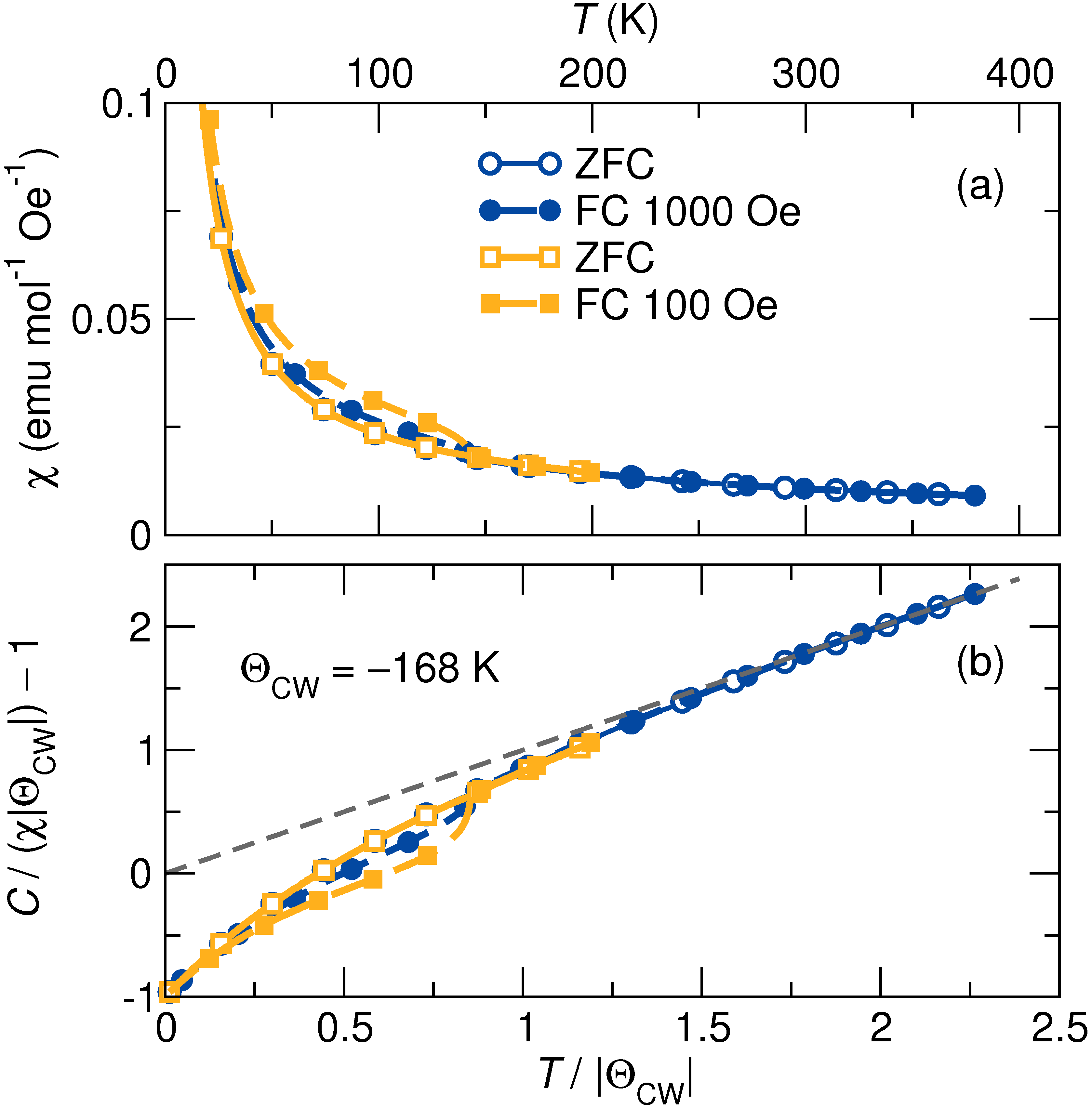} 
\caption{(a) Antiferromagnetic ordering of Ru$^{4+}$ 4d spins leads to a
small cusp in the ZFC magnetic susceptibility at 143\,K. Nd$^{3+}$ 4f
spins remain paramagnetic, and cause the upturn in susceptibility at lower
temperatures. The separation of the ZFC-FC susceptibility is tentatively attributed to
spin canting, which leads to weak ferromagnetism. (b) Scaled inverse 
susceptibility as a function of scaled temperature, as described in the text. 
The dashed line represents ideal Curie-Weiss behaviour, and the negative 
deviation in \nro\/ is due to short-range ferromagnetic interactions. 
The Curie-Weiss fit of the high-temperature data reveals a negative Weiss 
temperature, indicating that the dominant magnetic interactions 
are antiferromagnetic.}
\label{fig:zfcfc}
\end{figure}

\begin{figure}
\centering 
\includegraphics[width=0.5\textwidth]{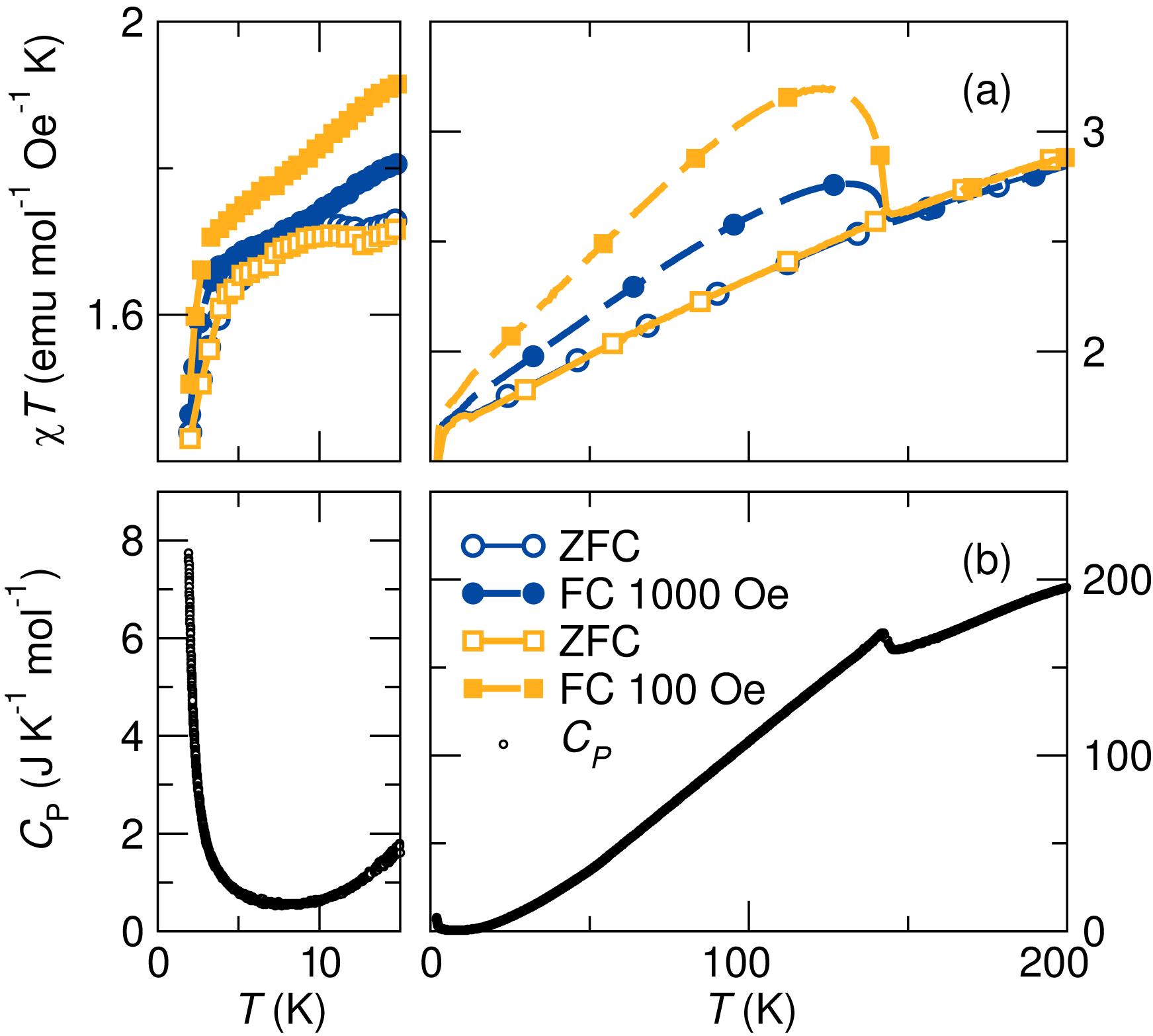} 
\caption{(a) The antiferromagnetic ordering seen in the DC magnetic
susceptibility at 143\,K is clearly visible in the $\chi T$ plot, and
corresponds closely with (b) the single anomaly in the heat capacity with a
maximum at 142\,K. The increase in heat capacity with decreasing temperature
at $T\leq 5$\,K is attributed to ordering of the Nd$^{3+}$ 4f spins.}
\label{fig:heatcapacity}
\end{figure}

Zero-field cooled (ZFC) and field cooled (FC) measurements of the magnetic
susceptibility show magnetic ordering of \nro\/ at $T_\text{N}=143$\,K 
(figure\,\ref{fig:zfcfc}). The higher-temperature region (340\,K to 380\,K) of 
the susceptibility data was fit to the Curie-Weiss (CW) equation,  
$\chi$ = $C/(T - \Theta_{\text{CW}})$. The effective moment was 
extracted using the relationship $\mu_{\text{eff}}^2 = 3C k_{\text{B}} / N$, while the 
estimated spin-only and unquenched moments of \nro\/ were calculated using 
$\mu_{\text{eff}}^2 = 2\mu_{\text{Ru}}^2+2\mu_{\text{Nd}}^2$. 
The determined effective paramagnetic moment was $\mu_{\text{eff}}$ = 6.36\,$\mu_{\text{B}}$ 
per \nro\/ formula unit and the Weiss temperature 
was $\Theta_{\text{CW}}=-168\,K$. The $\mu_{\text{eff}}$ is close to the  
calculated spin-only value of $\mu_{S}$ = 6.8\,$\mu_{\text{B}}$, and significantly less
than the calculated unquenched $\mu_{L+S}$ of 11.9\,$\mu_{\text{B}}$.  
Curie-Weiss analysis reveals a
negative $\Theta_{\text{CW}}$, indicating that the dominant magnetic interactions
are antiferromagnetic. However, it is apparent from the first derivative (not
shown) that the inverse susceptibility remains mildly non-linear as a
function of temperature, indicating that \nro\/ does not display pure
Curie-Weiss paramagnetism at these higher temperatures. Consequently, the
determined Weiss temperature and $\mu_{\text{eff}}$ should be treated only qualitatively. 

Rearranging the Curie-Weiss equation allows the scaled inverse susceptibility 
$C/(\chi|\Theta_{\text{CW}}|)-1$ to be displayed as a function of $T/|\Theta_{\text{CW}}|$ and 
provides a convenient way to visualize deviations from ideal Curie-Weiss 
paramagnetism [figure \ref{fig:zfcfc}(b)] \cite{Melot2009JPCM}. 
These deviations from Curie-Weiss behaviour are due to short-range interactions, 
and the negative deviation seen in \nro\/ arises from uncompensated moments. 
Additionally, the plot provides a convenient
method to visualize magnetic frustration (the frustration index 
$f$ = $\Theta_{\text{CW}}/T_\text{N}$) \cite{Ramirez1994ARMS}. Moderately frustrated 
systems tend to have $f \geq3$ \cite{Ramirez1994ARMS}, whereas for \nro\/ $f=1.2$, 
indicating the antiferromagnetic ordering is not strongly frustrated.

Previous studies have shown that the transition at 143\,K is the result of 
antiferromagnetic ordering of Ru$^{4+}$ 4d spins, as an
analogous transition is observed in Y$_2$Ru$_2$O$_7$, where there are no $f$
electrons at the \textit{A} site \cite{Gardner2010RMP,Ito2001JPCS}. The
negative Weiss temperature indicates the dominant magnetic interactions are
antiferromagnetic; however, the Nd$^{3+}$ 4f spins remain paramagnetic,
and cause the upturn in susceptibility at lower temperatures. Additionally,
the history-dependence of the ZFC and FC measurements and the increased
susceptibility at lower applied fields suggest there are weak uncompensated 
(\textit{i.e.} ferromagnetic)
moments. Owing to the dominant paramagnetic component of the susceptibility and
the normalization over the applied magnetic field, the weak ferromagnetic
interaction saturates at lower applied fields and its contribution appears
smaller at higher applied fields in the field-normalized susceptibility 
[figure \,\ref{fig:zfcfc}(a)]. 
Meanwhile, the ZFC-FC bifurcation of the magnetic susceptibility is
consistent with weak ferromagnetism arising from spin-canting, and is
described later in further detail. Below 10\,K, there is another transition
due to antiferromagnetic ordering that can be clearly seen in the $\chi T$
plot [figure\,\ref{fig:heatcapacity}(a)]. There is also a corresponding
increase in the specific heat that is attributed to the tail of a Schottky
anomaly, likely resulting from the ordering of the Nd 4f spins
\cite{Matsuhira2007JPSJ,Freamat2005PRB,Kittel2005}. The rare-earth
4f spins are expected to interact much less strongly than Ru 4d spins 
because the 4f-orbitals are contracted \cite{Blundell2001}. 
Consequently, Nd 4f spins are decoupled from the Ru 4d spins and order 
at much lower temperature \cite{Ito2000JPSJ}. 

\begin{figure}
\centering
\includegraphics[width=0.5\textwidth]{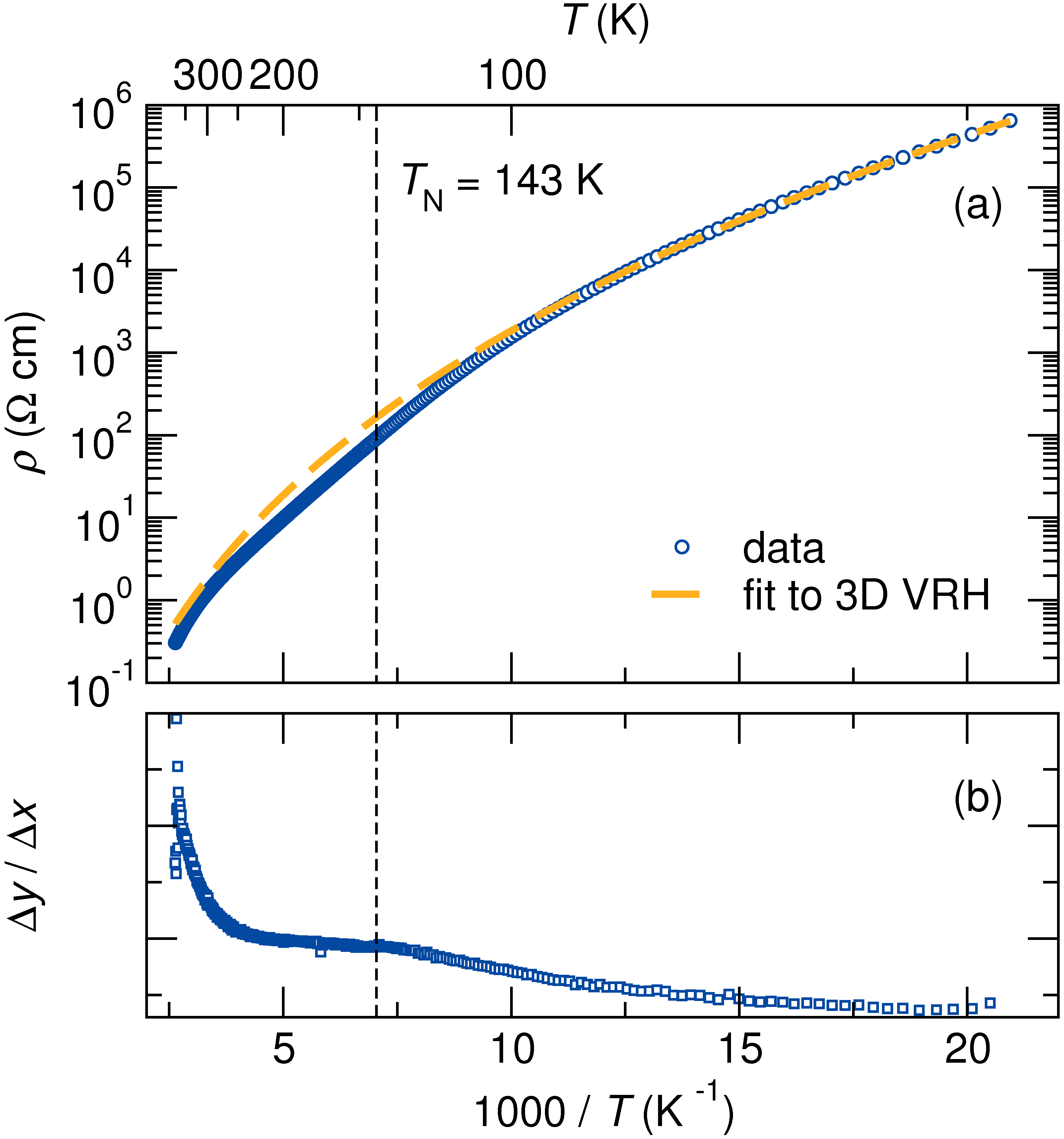}
\caption{(a) Low-temperature electrical resistivity follows a Mott 3D
variable-range hopping model, but deviates at the magnetic ordering
temperature ($T_\text{N}=143$\,K). Magnetic ordering of the Ru 4d conduction electrons
leads to a change in the electrical transport, as shown in the derivative (b).}
\label{fig:resistivity}
\end{figure}

The antiferromagnetic ordering of Ru 4d spins at $T_\text{N}=143$\,K causes notable effects 
in other measurements as well. The specific heat contains a corresponding $\lambda$-type
anomaly with a maximum at 142\,K [figure\,\ref{fig:heatcapacity}(b)], consistent with
a second-order phase transition.
Additionally, the electrical resistivity near the magnetic ordering
temperature displays anomalous behaviour. An Arrhenius-style plot shows a
change in slope at the magnetic ordering temperature 
(figure\,\ref{fig:resistivity}). Electrical conduction in \aro\/ materials 
involves Ru 4d states, so it should come as no surprise that magnetic 
ordering of Ru 4d spins causes a marked change in electrical transport due 
to changes in scattering, as shown in the derivative of 
figure\,\ref{fig:resistivity}. Low-temperature electrical resistivity follows 
a 3D variable-range hopping model with $\rho(T)$ = 
$\rho_0\text{exp}(\frac{T_0}{T})^{\frac14}$ 
\cite{Mott1968JNS,Rosenbaum1997JPCM}, with a change in the hopping barrier at
the magnetic ordering temperature ($T_\text{N}=143$\,K).

There has been considerable confusion in the literature about the nature of
the ordering of the Ru 4d spins in \nro\/ and analogous Ru-pyrochlores, in
part due to the unusual field-dependent hysteresis present between the ZFC
and FC susceptibility measurements, and also due to the many types of
disorder and exotic phenomena that sometimes accompany geometric frustration
in magnetic pyrochlores. In particular, several reports have mentioned that
the glassy nature of the ordering of Ru 4d spins is evident in bulk
magnetic susceptibility measurements
\cite{Taira2000JSSC,Ito2000JPSJ,Ito2001JPCS,Kmiec2006PRB,Gurgul2007PRB}.
However, neutron diffraction experiments below the ordering temperature show
that Ru spins order with a long correlation length 
\cite{Ito2000JPSJ,Ito2001JPCS}. This is incompatible with a
glassy-state, where there is no long-range magnetic order. This is not the
only signature of a spin-glass that is violated by experimental evidence.
Notably, the temperature-dependence of the specific heat should vary smoothly
near the magnetic ordering temperature \cite{Fischer1985PSSB}, a stark
contrast with the $\lambda$-type anomaly at 143\,K, shown in 
figure\,\ref{fig:heatcapacity}(b).

\begin{figure}
\centering 
\includegraphics[width=0.5\textwidth]{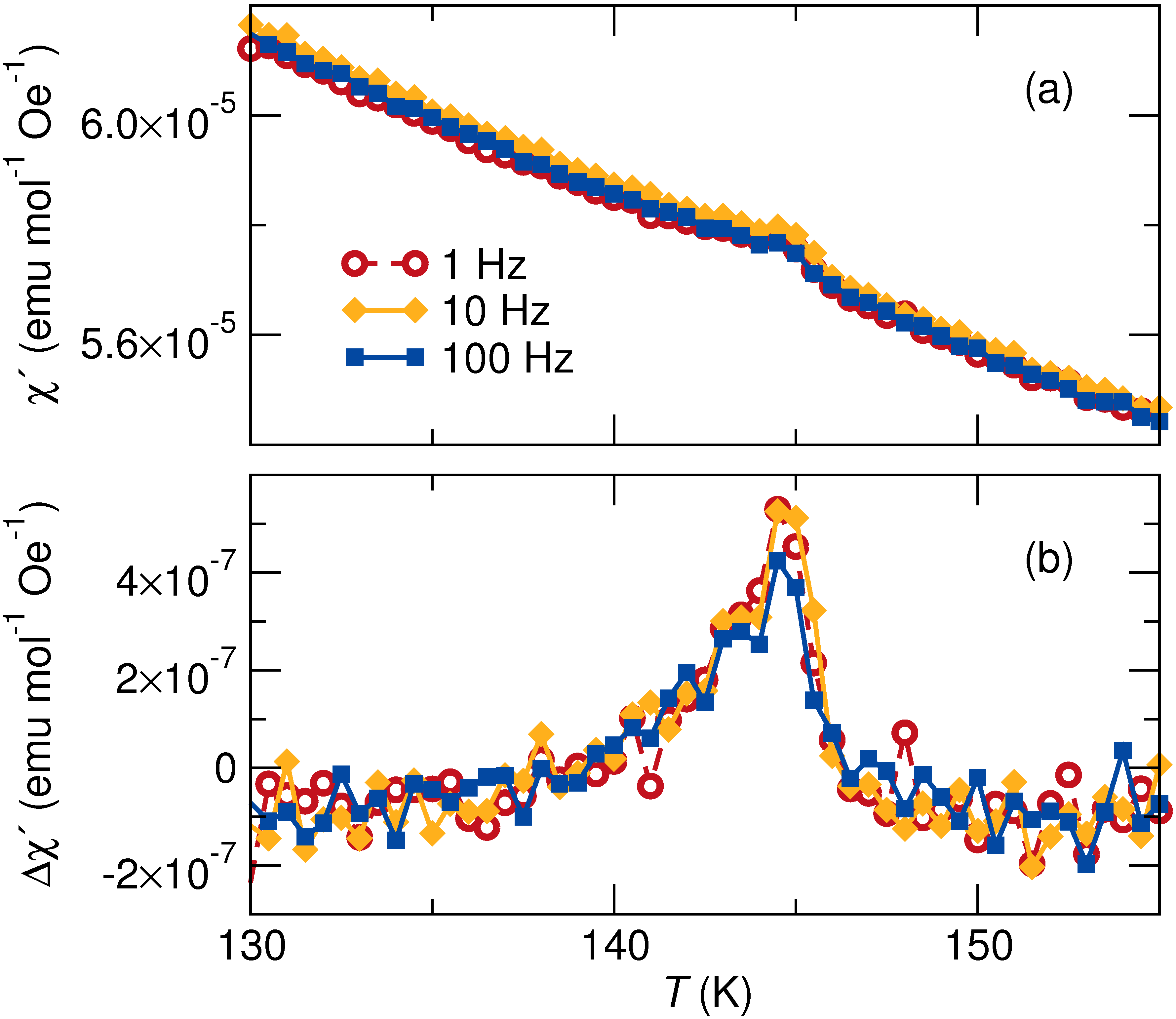} 
\caption{(a) The real component of the AC susceptibility shows a small peak
due to the antiferromagnetic ordering of Ru 4d spins. (b)
Background-subtracted data show the transition temperature is
frequency-independent, which indicates the antiferromagnetic ordering of
Ru 4d spins is not glassy, and that \nro\/ is not a spin-glass.}
\label{fig:ac_susceptibility}
\end{figure}

AC magnetic susceptibility measurements were also conducted, as another
signature of a spin-glass is a magnetic ordering temperature that varies with
the frequency of the applied magnetic field 
\cite{Fischer1985PSSB,Binder1986RMP}. 
The real, in-phase component of the AC susceptibility 
$\chi^\prime$, is shown in figure\,\ref{fig:ac_susceptibility}(a),
while the imaginary, out of phase component ($\chi^{\prime\prime}$) of the 
complex susceptibility was below the instrumental detection limit, and
is thus not presented. Owing to the paramagnetic response of the Nd$^{3+}$
that dominates the signal, the paramagnetic background was fit to the
Curie-Weiss equation and was subtracted to yield the antiferromagnetic 
component of the signal [figure\,\ref{fig:ac_susceptibility}(b)].
The antiferromagnetic ordering temperature associated with the Ru 4d spins is 
independent of frequency, which is inconsistent with canonical spin-glass 
behaviour \cite{Fischer1985PSSB,Binder1986RMP}.

A spin-glass, by definition, must lack long-range order and display a
frequency-dependent peak in the susceptibility
\cite{Fischer1985PSSB,Binder1986RMP}; neither condition is satisfied in
\nro\/. Additionally, given that the rare-earth 4f-spins are decoupled from
the Ru 4d-spins, the nature of the magnetic ordering is not expected to
differ in the \aro\/ (\textit{A}\,=\,Pr, \ldots Lu, Y) series  as the identity
of the rare-earth atom changes, so it is also unlikely that other members are
spin-glasses, contrary to previous reports
\cite{Taira2000JSSC,Ito2000JPSJ,Ito2001JPCS,Kmiec2006PRB,Gurgul2007PRB}. 

Adding to the confusion regarding the magnetic behaviour of \aro\/
(\textit{A} = Pr, \ldots Lu, Y), previous studies of \nro\/ have claimed there
is ferromagnetic ordering at 20\,K, and reported a hysteresis in the
low-temperature field-dependent magnetization \cite{Taira1999JPCM}, neither
of which were observed here in samples of $>$99\% purity. 
Additionally, a later study of the specific heat of \nro\/ showed multiple 
anomalies at 130\,K and 20\,K, in addition to the anomaly observed
here at 142\,K \cite{Taira2000JSSC}. However, the ferromagnetic ordering at
20\,K and the other specific heat anomalies are features of \nnro\/, a
secondary phase that is easily formed during preparation of \nro\/ 
and to which we attribute the
ferromagnetic ordering and other specific heat anomalies. Extensive study of
\nnro\/ by neutron diffraction, magnetization, and specific heat measurements
has shown the material undergoes a transition with a peak at 19\,K due to
ordering of both Ru$^{5+}$ and Nd$^{3+}$ spins \cite{Harada2001JPCM}.
Additionally, there is a peak in the specific heat at 130\,K, corresponding
to a first-order structural phase transition. Although preparation of \aro\/
(\textit{A} = Pr, \ldots Lu, Y) by ceramic reaction appears
straightforward, care must be taken at high temperatures in air to prevent
slow decomposition of the product. To demonstrate this, \nro\/ was annealed
in air at 1040\degree C and 1060\degree C for two weeks, and led to the
formation of \nnro\/ as a dominant, or single phase. In the literature,
preparation of \nro\/ appears to consistently lead to appreciable amounts of
\nnro\/ as a secondary phase, with some studies showing up to 10\,mol\%
\cite{Ito2000JPSJ}. Due to the similar properties and chemistry of analogous
\aro\/ systems, it is not surprising to see similar features near 20\,K in
the magnetic susceptibility and the specific heat (\textit{e.g.},
ferromagnetic ordering and a $\lambda$-type anomaly)\cite{Taira2000JSSC}, 
as these features may originate from an \aaro\/ secondary phase. 

\subsection{Electrical transport in Ru pyrochlores}

The nature of the electrical transport behaviour in ruthenium pyrochlores
(\aro\/) has been of considerable interest, as the nature of the
\textit{A}-site ion dictates whether the material will be insulating
(\textit{e.g.}, \textit{A} = Pr, \ldots Lu, Y) or metallic (\textit{e.g.},
\textit{A}\,=\,Tl, Pb, Bi)\cite{Li2003CM,Field2000JSSC,Takeda1998JSSC,Ishii2000JPSJ,Tachibana2006PRB}. 
The most intuitive model has sought to explain the different behaviour strictly 
with the $A$-site ionic radius, as increasing the $A$-site radius increases 
the Ru--O--Ru bond angle. Electrical conduction
takes place in the Ru$_2$O$_6$ network through overlap of Ru 4d and
O 2p orbitals, so a larger Ru--O--Ru bond angle increases this
overlap, the bandwidth, and the electrical conductivity
\cite{Lee1997JSSC,Li2003CM,Ishii2000JPSJ}.

Unfortunately, testing this relationship is not straightforward, as no
lanthanide leads to metallic behaviour. Although the La$^{3+}$ radius is
similar to Bi$^{3+}$ and should thus be large enough to induce metallic
behaviour, La$_2$Ru$_2$O$_7$ is outside the pyrochlore stability-field
\cite{Subramanian1983PSSC,Gardner2010RMP,Cai2011JMC}. Several studies
interested in the electrical conductivity of these systems have examined
solid solutions with Bi or Pb to overcome this hurdle and increase the
average ionic radius to the point at which the system becomes metallic
\cite{Li2003CM,Yamamoto1994JSSC,Kennedy1996JSSC,Kanno1993JSSC,Field2000JSSC,Kobayashi1995JSSC}.
However, it is not appropriate to compare rare-earth Ru pyrochlores with those 
containing Pb or Tl (\textit{e.g.} Pb$_2$Ru$_2$O$_{6.5}$,
Tl$_2$Ru$_2$O$_{7-y}$), as these contain either \textit{A}-site ions of
different formal charge in the case of Pb$^{2+}$, or have contributions 
from overlap between empty 6s states and filled states, as in the case of
Tl$^{3+}$. Even an isovalent substitution of $Ln^{3+}$ by Bi$^{3+}$ may not 
allow straightforward comparison; structural studies of \bro\/ have shown that
Bi$^{3+}$ atoms are off-centered due to the stereochemically-active $6s^2$
lone-pair \cite{Shoemaker2011PRB,Avdeev2002JSSC}. Indeed, Kennedy and Vogt,
among others, point out that, while several structural or physical parameters
of insulating \aro\/ (\textit{A}\,=\,Pr, \ldots Lu, Y) pyrochlores tend to
follow a simple trend (\textit{e.g.}, a linear variation in oxygen
positional parameter and Ru--O bond distance as a function of lattice
parameter), the behaviour of metallic ruthenium pyrochlores is aberrant
\cite{Lee1997JSSC,Kennedy1996JSSC}. 

\begin{figure}
\centering 
\includegraphics[width=0.5\textwidth]{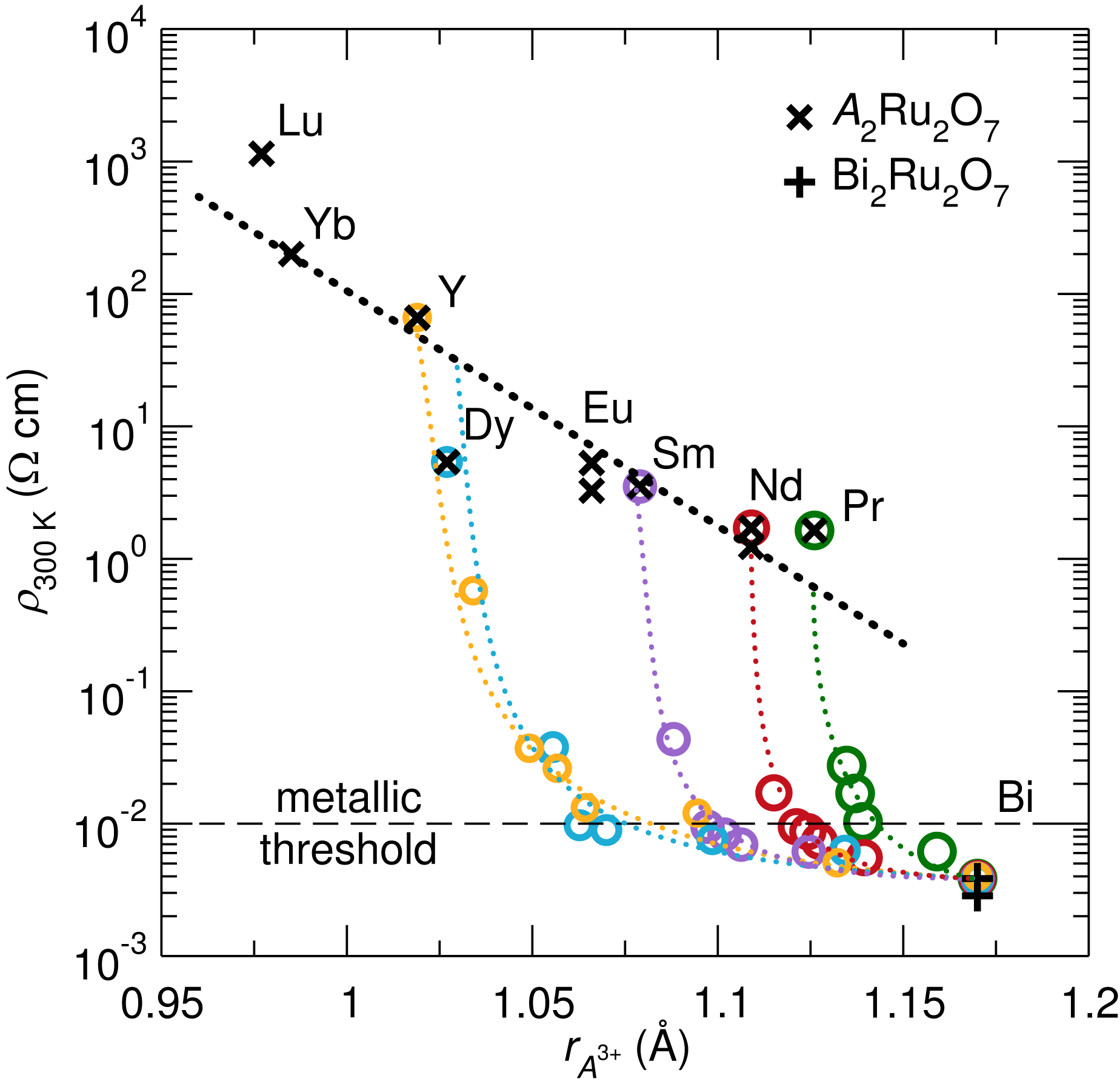} 
\caption{Room-temperature electrical resistivity of \abro\/ 
(\textit{A} = Pr, \ldots Lu, Y) solid solutions decrease smoothly with 
increasing average cation radius. Introduction of Bi drastically changes
the behaviour (circles). Dotted lines are guides to the eye, and
the thin dashed line at $\rho=0.01$\,$\Omega$\,cm represents the Mott 
minimum metallic conductivity at room temperature. Ionic radii of 8-coordinate 
3+ cations were taken from Shannon \protect\cite{Shannon1976ACSA}. Values of electrical 
resistivity were taken from references \protect\cite{Subramanian1983PSSC,Li2003CM,Yamamoto1994JSSC,Kanno1993JSSC,Bouchard1971MRB,Munoz-Perez2012JAP,Beyerlein1988JSSC,Sleight1972NSP}.}
\label{fig:size_effects}
\end{figure}

To examine the influence of the \textit{A}-site ionic radius on the electrical
transport, measurements of the electrical resistivity at 300\,K for several 
Ru pyrochlore
systems have been gathered and are presented in figure\,\ref{fig:size_effects}.
Although examining the relationship between the Ru--O--Ru bond angle and
electrical resistivity might be more direct, the lack of reliable structural
data makes the number of available systems less informative. As mentioned
earlier in detail, Kennedy and Vogt demonstrated the use of lab XRD has led to
inaccurate determination of the O atomic position\cite{Kennedy1996JSSC}. 
On the other hand, the ionic radius is independent of the
reported crystallographic data, so it is used as the abscissa in this case.
Additionally, when examining these systems and structure-property
relationships, it is common to show parameters plotted \textit{vs.} lattice parameter.
However, the variation in lattice parameter is strongly influenced by changes in
the nature of bonding, as in the case of metallic \bro\/. For example, even
though 8-coordinate Bi$^{3+}$ has a larger ionic radius than Nd$^{3+}$,
($r_{Bi^{3+}}=1.17$\,\AA\/, $r_{Nd^{3+}}=1.109$\,\AA\/)
\cite{Shannon1976ACSA}, \bro\/ has a smaller unit cell than \nro\/
\cite{Yamamoto1994JSSC}.

Casual examination of figure\,\ref{fig:size_effects} suggests a monotonic
decrease in the room-temperature resistivity as the ionic radius of the 
\textit{A} site increases, when only the rare-earth-containing pyrochlores 
(\aro\/, \textit{A}\,=\,Pr, \ldots Lu, Y) are considered. However, 
substitution with Bi dramatically changes the behaviour. The
difference in electrical resistivity is truly striking, and suggests that there is more
than the effect of ionic radius when Bi is incorporated into the material. We
argue that local distortion of Bi centres in Ru pyrochlores is responsible
for the metallic behaviour and the distinct behaviour of \bro\/ and \abro\/
when contrasted with Bi-free samples. Other researchers
have pointed to the presence of additional oxygen vacancies (other than the
ordered vacancy dictated by the pyrochlore structure) and decreased
Ru--O bond distance that accompany metallic behaviour
\cite{Li2003CM,Field2000JSSC}, though these are a \textit{result},
rather than the cause, of metallic bonding.

\subsection{High-temperature thermoelectric properties}

\begin{figure}
\centering 
\includegraphics[width=0.5\textwidth]{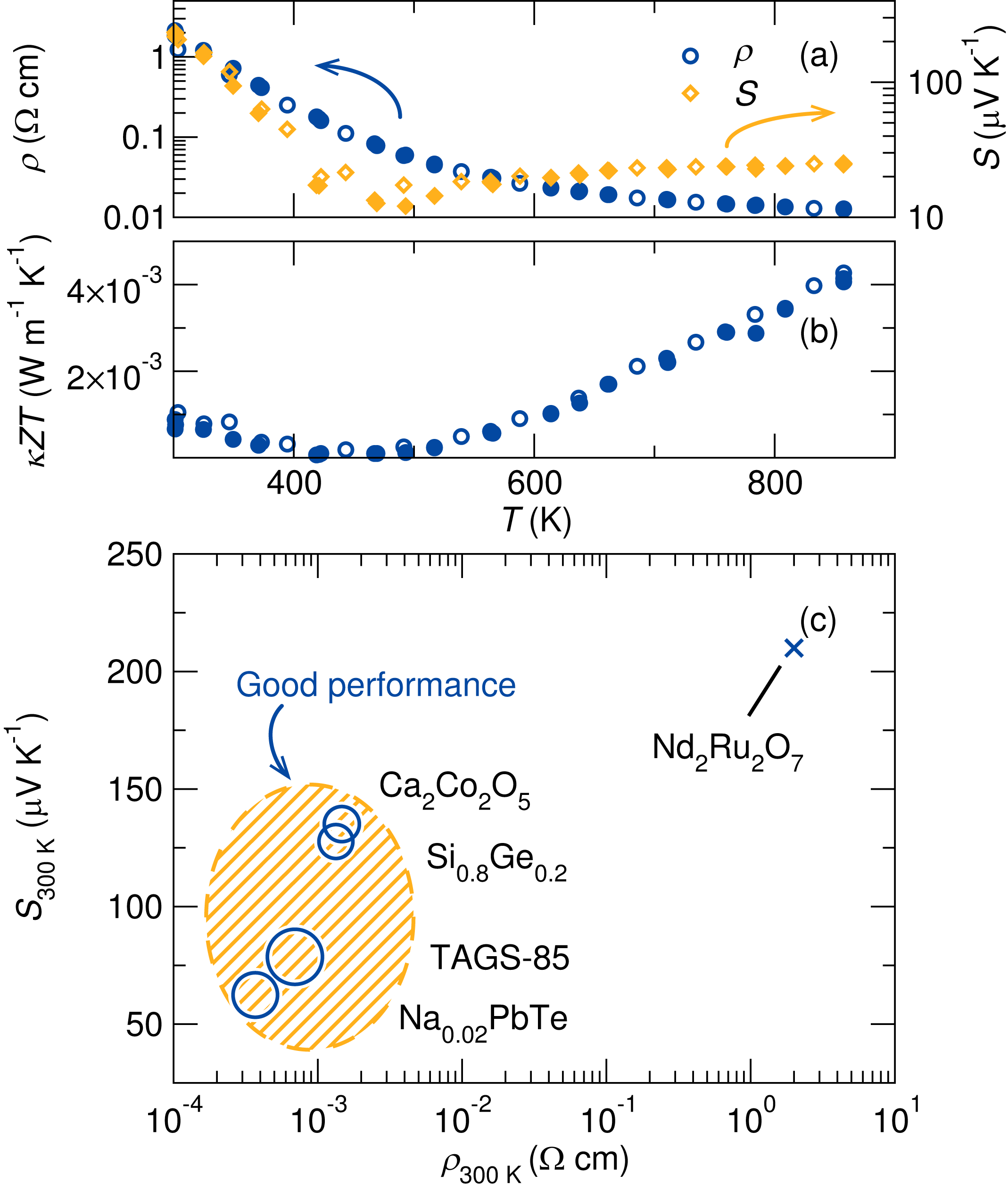} 
\caption{(a-b) High-temperature thermoelectric properties demonstrate that
\nro\/ has poor thermoelectric performance from 300\,K to 900\,K. Three heating
and cooling cycles were performed to ensure the stability of the sample; the
first cycle is represented by hollow symbols, subsequent cycles by filled
symbols. (c) Comparison with high-temperature $p$-type thermoelectrics
reveals that high-performance materials are clustered in one region of the plot,
while the performance of \nro\/ is limited by high resistivity.
This method of visualization can be used as a rapid 
screening tool to determine whether a material is likely to have
reasonable thermoelectric performance. The radius of the circle is $\kappa$\textit{ZT} 
 at 700\,K. Since $\kappa$\textit{ZT} of \nro\/ is very small,
it is represented by $\times$. Values are presented for Ca$_2$Co$_2$O$_5$ 
 \cite{Funahashi2000JJAPP2}, Si$_{0.8}$Ge$_{0.2}$ \cite{Vining1991JAP}, 
(AgSbTe$_2$)$_{0.15}$(GeTe)$_{0.85}$ (TAGS-85) \cite{Rowe1995}, and 
Na$_{0.02}$PbTe \cite{Pei2011N}.} 
\label{fig:ZEM}
\end{figure}

The electrical transport properties of rare-earth 
\aro\/ (\textit{A}\,=\,Pr\, \ldots Lu, Y) members are near the onset of metallic behaviour. 
This unique position is a good place to examine thermoelectric properties, where there is a balance
between a high Seebeck coefficient and low electrical resistivity. Electrical
resistivity and Seebeck coefficient of \nro\/ are presented from 
300\,K to 900\,K (figure\,\ref{fig:ZEM}). The thermoelectric properties of 
\nro\/ do not change over three measurement cycles, despite the low oxygen 
partial pressure and high temperature during measurements. 
The room-temperature electrical resistivity is 2.0\,$\Omega$\,cm, and decreases 
with increasing temperature, as expected for a non-metal. The room-temperature 
Seebeck is promising ($\approx$220\,$\mu$V/K), but quickly decreases 
with increasing temperature, and  $\approx$20\,$\mu$V/K above 420\,K.
$\kappa$\textit{ZT} is also presented [figure\,\ref{fig:ZEM}(b)], as a proxy for the 
thermoelectric figure of merit, and is several orders of magnitude too small for 
\nro\/ to be a competitive thermoelectric material.
The thermoelectric figure of merit, \textit{ZT}, is given by 
$ZT = S^2T/(\rho \kappa)$ and is a function of the Seebeck coefficient \textit{S}, 
electrical resistivity $\rho$, thermal conductivity $\kappa$, and temperature 
\textit{T}. Many materials exhibit a thermal conductivity between 
1\,W\,m$^{-1}$K$^{-1}$ and 10\,W\,m$^{-1}$K$^{-1}$, so $\kappa$\textit{ZT} is a useful 
proxy to compare the electrical performance of thermoelectric 
materials and estimate \textit{ZT} within an order of magnitude.

Following our investigation of \nro\/ and learning of its low $\kappa$\textit{ZT} 
at high temperatures, we realized that a simple analytical method for estimating 
the competitiveness of a thermoelectric at high temperatures would be useful for 
identifying new candidate materials. Such a method would be especially helpful 
if it did not require high-temperature measurements, as this would save time and resources.
 Many materials are currently being screened
for high-temperature thermoelectric performance, but measurements at elevated
temperatures require specialized instrumentation that is not widely
available. Room-temperature Seebeck measurements can be performed quickly and
with less sophisticated equipment. Additionally, low-temperature
(\textit{i.e.}, \textit{T}\,$\leq$\,300\,K) Seebeck and electrical resistivity data are
available in the literature for many materials, and could be used to quickly
eliminate materials that are likely to have poor thermoelectric performance.

Towards this end, a new type of plot is presented in figure\,\ref{fig:ZEM}(c),
where the room-temperature Seebeck ($S_{300\,K}$) is plotted versus
the room-temperature electrical resistivity ($\rho_{300\,K}$). The
radius of a data point represents the magnitude of $\kappa$\textit{ZT} at an arbitrary temperature
that is common among the data points, and allows easy comparison of the expected 
thermoelectric performance at that temperature. When examining \nro\/ and its relation
to other high-temperature $p$-type materials at 700\,K using the aforementioned survey 
plot [figure\,\ref{fig:ZEM}(c)], high-performance materials are clustered
in one region of the plot (filled area) while \nro\/ is isolated. In particular, the location of 
\nro\/ indicates the room-temperature electrical resistivity is too high.
Because the $\kappa$\textit{ZT} of \nro\/ is very small and leads to a vanishingly small point, 
it is represented by the symbol $\times$ in figure\,\ref{fig:ZEM}(c). 

This type of analysis is particularly useful in establishing that the performance of a particular
class of materials may be ineffectual due to a key property being outside the useful range.
The choice of axes is similar to a Jonker plot, in which the Seebeck 
coefficient is plotted \textit{vs.} electrical conductivity. Jonker plots 
are used to examine the effect of changing the carrier concentration of 
a single material rather than looking at a field of candidate thermoelectics
\cite{Jonker1968PRR,Bak2011JPCC}. Also, while Zhu \textit{et al.} have shown
that Jonker plots could be used to to estimate the peak thermoelectric
power factor ($S^{2} \rho^{-1}$) of a material \cite{Zhu2011JACS}, 
other analyses are involved, and the approach is distinct from the one employed here.

\section{Conclusion}

Pyrochlore \nro\/ has been prepared and examined using a combination of structural, magnetic, 
and electrical and thermal transport studies. Some substitutional disorder on the \textit{A}-site 
is proposed from the structural studies, but is not anticipated to strongly influence 
the physical properties. The magnetic behaviour of \nro\/ has been clarified through a combination of 
DC and AC magnetic measurements, and heat capacity studies. Despite the potential for geometric 
frustration of magnetism in the pyrochlore structure-type, we find no such exotic behaviour or 
glassiness in \nro\/ and instead tentatively suggest it is a canted antiferromagnet that displays weak 
ferromagnetism. When the electrical transport properties are regarded in light of published 
data on rare-earth substituted \aro\/ pyrochlores, it is clear that ionic radius plays a key 
role in determining electrical behaviour. However, the metallic electrical properties that accompany
incorporation of Bi$^{3+}$ on the \textit{A}-site lies outside this description, and we 
suggest Bi$^{3+}$ off-centering may drive this anomalous behaviour.
High temperature measurements of the thermoelectric properties indicate that 
\nro\/ has excessively high electrical resistivity for it to be a useful 
thermoelectric, despite displaying a promising Seebeck coefficient at room temperature.
We propose a modified version of the Jonker plot as a powerful tool to screen
candidate thermoelectric materials. We find that it is particularly useful in 
establishing that a particular materials class may be ineffectual due to a key 
property being outside the useful range.

\ack
We thank the National Science Foundation for support of this research through
NSF-DMR 1121053. Prof.\,Carlos Levi is thanked for his involvement and guidance. 
Dr.\,Badri Shyam and Dr.\,Daniel Shoemaker are thanked for 
collecting the total scattering data at beamline 11-ID-B at the APS, and 
Dr. Brian Toby is thanked for helpful discussions on powder diffraction. 
MWG thanks the Natural Sciences and Engineering Council of
Canada and the US Department of State for support through a NSERC
Postgraduate Scholarship and an International Fulbright Science \& Technology
Award, respectively. PTB is supported by the NSF Graduate Research Fellowship Program.
CSB is a recipient of the Feodor Lynen Research Fellowship 
supported by the Alexander von Humboldt foundation.
The research carried out here made extensive use of shared
experimental facilities of the Materials Research Laboratory: an NSF MRSEC,
supported by NSF-DMR 1121053. The MRL is a member of the the NSF-supported
Materials Research Facilities Network (www.mrfn.org). Use of data from beamlines
11-BM and 11-ID-B at the Advanced Photon Source is supported by the Department
of Energy, Office of Science, Office of Basic Energy Sciences, under Contract
No. DE-AC02-06CH11357. Neutron research facilities used in this work were
provided by the National Institute of Standards and Technology, US Department
of Commerce.

\section*{References}

\bibliographystyle{iopart-num}
\bibliography{Nd2Ru2O7}

\end{document}